\documentclass[journal]{IEEEtran}

\usepackage{xcolor,soul,framed} 

\colorlet{shadecolor}{yellow}
\usepackage[pdftex]{graphicx}
\graphicspath{{../pdf/}{../jpeg/}}
\DeclareGraphicsExtensions{.pdf,.jpeg,.png}

\usepackage[cmex10]{amsmath}
\usepackage{array}
\usepackage{amsmath}
\usepackage{amssymb}
\usepackage{amsfonts}
\usepackage{mdwmath}
\usepackage{mdwtab}
\usepackage{placeins}
\usepackage{eqparbox}
\usepackage{url}
\usepackage{caption} 
\usepackage{booktabs} 
\usepackage{textcomp}

\begin{document}
\bstctlcite{IEEEexample:BSTcontrol}
    \title{ChordFormer: A Conformer-Based Architecture for Large-Vocabulary Audio Chord Recognition}
  \author{Muhammad Waseem~Akram,
      Stefano~Dettori, 
      Valentina~Colla, 
      Giorgio~Carlo~Buttazzo

  \thanks{Muhamamd Waseem Akram is with Department of Excellence in Robotics and AI, Sant'Anna School of Advanced Studies, Itay (e-mail: muhammadwaseem.akram@santannapisa.it).}
  \thanks{Stefano Dettor is with Department of Excellence in Robotics and AI, Sant'Anna School of Advanced Studies, Italy (e-mail: stefano.dettori@santannapisa.it).}%
  \thanks{Valentina Colla is with Department of Excellence in Robotics and AI, Sant'Anna School of Advanced Studies, Italy (e-mail: valentina.colla@santannapisa.it).}
  \thanks{Giorgio Carlo Buttazzo is with Department of Excellence in Robotics and AI, Sant'Anna School of Advanced Studies, Italy (e-mail: giorgio.buttazzo@santannapisa.it).}
}


\maketitle

\begin{abstract}
Chord recognition serves as a critical task in music information retrieval due to the abstract and descriptive nature of chords in music analysis. While audio chord recognition systems have achieved significant accuracy for small vocabularies (e.g., major/minor chords), large-vocabulary chord recognition remains a challenging problem. This complexity also arises from the inherent long-tail distribution of chords, where rare chord types are underrepresented in most datasets, leading to insufficient training samples. Effective chord recognition requires leveraging contextual information from audio sequences, yet existing models, such as combinations of convolutional neural networks, bidirectional long short-term memory networks, and bidirectional transformers, face limitations in capturing long-term dependencies and exhibit suboptimal performance on large-vocabulary chord recognition tasks. This work proposes ChordFormer, a novel conformer-based architecture designed to tackle structural chord recognition (e.g., triads, bass, sevenths) for large vocabularies. ChordFormer leverages conformer blocks that integrate convolutional neural networks with transformers, thus enabling the model to capture both local patterns and global dependencies effectively. By addressing challenges such as class imbalance through a reweighted loss function and structured chord representations, ChordFormer outperforms state-of-the-art models, achieving a 2\% improvement in frame-wise accuracy and a 6\% increase in class-wise accuracy on large-vocabulary chord datasets. Furthermore, ChordFormer excels in handling class imbalance, providing robust and balanced recognition across chord types. This approach bridges the gap between theoretical music knowledge and practical applications, advancing the field of large-vocabulary chord recognition.

\end{abstract}


%
\IEEEpeerreviewmaketitle


\section{Introduction}

\IEEEPARstart{A}{utomatic} chord recognition (ACR) forms the foundation of music information retrieval, acting as a key tool for understanding and analyzing the harmonic structure of music. ACR systems aim at decoding audio recordings into sequences of time-synchronized chord labels, enabling various applications such as automatic lead-sheet creation~\cite{duran2020transcribing}, music structure analysis~\cite{de2022measuring}, key classification~\cite{weiss2020local}, and cover song identification~\cite{du2023bytecover3}. These systems not only assist musicians, composers, and educators but also facilitate a deeper understanding of musical content, aiding in the preservation and exploration of diverse musical traditions. Despite its transformative potential, ACR remains a challenging task due to the inherent complexities of musical chords, their temporal structures~\cite{pauwels2019acr}, and the uneven distribution of chord types in real-world music datasets~\cite{jiang2019large, rowe2021curriculum}.

Chords, as fundamental elements of harmony, are more than just combinations of notes~\cite{de2023choco}. They carry rich musical semantics and temporal relationships that are rooted in centuries of music theory and practice. These relationships are far from random, often following well-defined harmonic progressions that exhibit long-term dependencies. Modeling these dependencies is crucial for accurate chord recognition, but it is challenging task~\cite{pauwels2019acr}. Standard approaches, such as Hidden Markov Models (HMMs)~\cite{sheh2003chord,ueda2010hmm}, have historically relied on transition matrices to capture these temporal dynamics, but they struggle to extend beyond short-term dependencies due to their limited capacity to model complex and long-range relationships. Similarly, $n$-gram models~\cite{khadkevich2009use} and feedforward neural networks~\cite{korzeniowski2016feature} have been explored, but their effectiveness is limited by the relatively small harmonic context they can capture.

The advent of deep learning has brought significant advancements to ACR~\cite{sigtia2015audio, zhou2015chord, korzeniowski2016feature}, enabling models to learn from large datasets and capture intricate patterns in musical data. Architectures such as convolutional neural networks (CNNs)~\cite{jiang2019large,humphrey2012rethinking,korzeniowski2016fully,mcfee2017structured} and recurrent neural networks (RNNs)~\cite{sigtia2015audio,wu2018automatic,boulanger2013audio,deng2016hybrid,deng2017large} have been widely adopted, offering the ability to jointly model chord sequence consistency, chord duration, and other related features. However, these architectures are not without limitations. CNNs, for instance, are inherently designed to capture local patterns and struggle with long-term dependencies, while RNNs often face issues such as vanishing gradient, which hinder their ability to remember distant temporal information. Such limitations have spurred the exploration of attention mechanisms and transformer-based architectures, which have revolutionized sequence modeling in fields like natural language processing~\cite{vaswani2017attention, islam2024comprehensive}, speech recognition~\cite{gulati2020conformer}, and music generation~\cite{civit2022systematic}.

Transformers, characterized by their self-attention mechanisms, excel at capturing long-range dependencies and global context without relying on recurrence or convolution. Their ability to dynamically weigh the importance of different parts of an input sequence has proven transformative in various domains. In the context of ACR, such capabilities hold promise for addressing long-term chord dependencies, a challenge that has historically limited the performance of many systems~\cite{jiang2019large,sigtia2015audio, humphrey2012rethinking,korzeniowski2016fully,mcfee2017structured,wu2018automatic,boulanger2013audio,deng2016hybrid,deng2017large}. In addition to the challenges of modeling temporal dependencies, ACR faces significant hurdles related to chord vocabulary and class imbalance~\cite{jiang2019large, rowe2021curriculum}. 

Large-vocabulary chord recognition is inherently complex due to the vast number of chord qualities and the subtle differences between them. For instance, extended chords such as ninths, elevenths, and thirteenths often share overlapping chromatic features, making them difficult to distinguish. Moreover, the distribution of chord types in real-world datasets is highly imbalanced, with a small number of chord types dominating most of the musical content. This imbalance skews learning processes, causing models to overfit common chord types and underperform on rarer ones. Addressing chord imbalance is critical for developing robust and generalizable ACR systems.

To tackle these challenges, this paper introduces a new model, ChordFormer, for structural chord recognition based on conformer blocks~\cite{gulati2020conformer}. The conformer architecture combines convolutional layers with self-attention mechanisms and is particularly well-suited for this task, as it effectively balances the modeling of local and global dependencies. By leveraging such a hybrid design, our model achieves a nuanced understanding of chord sequences, overcoming the limitations of existing approaches. ChordFormer incorporates a reweighted loss function~\cite{jiang2019large}, ensuring that underrepresented chord classes receive an adequate attention during training. This approach mitigates the overemphasis on frequent chord types, promoting a more balanced learning across the chord vocabulary. Furthermore, ChordFormer adopts a structured representation of chord labels inspired by chord structure decomposition proposed by Jiang et al.~\cite{jiang2019large}. Unlike traditional flat classification methods, this representation captures the hierarchical and semantic relationships between chord qualities, enabling the model to better differentiate between subtle variations in chord structures. By integrating these innovations, our approach not only improves recognition accuracy, but also provides a more musically meaningful interpretation of chord labels.

In summary:
\begin{enumerate}
    \item A new Conformer-based architecture is proposed for a structural chord recognition, leveraging its hybrid design to capture both local and long-term dependencies in chord sequences.
    \item The pervasive issue of class imbalance is addressed through a reweighted loss function, which enhances the model’s performance on rare chord types.
    \item A structured representation of chord labels that aligns with musical theory, is applied to offer a more interpretable and effective approach to large-vocabulary chord recognition.
\end{enumerate}

The effectiveness of the proposed approach has been evaluated by conducting extensive experiments on a chord recognition dataset comprising 1,217 songs compiled by Humphrey and Bello~\cite{mcfee2017structured, humphrey2015four}. The achieved results demonstrate that the ChordFormer model outperforms state-of-the-art methods, including those employing large-vocabulary transcription and chord structure decomposition. It exhibits superior performance across various metrics, with notable improvements in recognizing rare and complex chord types.

The remainder of this paper is organized as follows. Section~\ref{rw} reviews related work in chord recognition. Section~\ref{arc} details the design and implementation of the ChordFormer model, including its reweighted loss function and structured label representation. Section~\ref{exp} presents the experimental setup and evaluation metrics, while the results are discussed in Section~\ref{res}. Finally, Section~\ref{conc} concludes the paper and outlines some directions for future research in this domain.

\section{Related work}\label{rw}

Automatic Chord Recognition (ACR) has undergone substantial evolution since Fujishima's groundbreaking work in 1999~\cite{takuya1999realtime}. This pioneering effort introduced a real-time system utilizing 12-dimensional chroma features and Hidden Markov Models (HMMs) to model chord sequences. Fujishima’s approach transitioned ACR from handcrafted feature-based methodologies into data-driven paradigms powered by advancements in machine learning and deep learning~\cite{mcvicar2014automatic}.

Initial ACR methodologies focused on feature extraction and chord sequence decoding. Techniques such as chroma vectors and tonal centroid representations, including the Tonnetz~\cite{humphrey2012learning}, were instrumental in capturing harmonic content. Probabilistic models like Gaussian Mixture Models (GMMs)~\cite{khadkevich2013reassigned} and HMMs~\cite{ueda2010hmm,sheh2003chord, khadkevich2009use} served as the backbone for sequence decoding. Despite their utility, these approaches faced limitations in modeling complex harmonic progressions and handling large vocabularies. Incremental improvements~\cite{ni2012end, mauch2010approximate}, such as higher-order HMMs, alleviated some issues but were restricted by the inadequacies of manually designed features.

The introduction of deep learning marked a paradigm shift in ACR~\cite{korzeniowski2016feature, sigtia2015audio, zhou2015chord}. Convolutional neural networks (CNNs) and recurrent neural networks (RNNs)~\cite{jiang2019large,humphrey2012rethinking,mcfee2017structured,wu2018automatic,boulanger2013audio,deng2016hybrid,deng2017large} became dominant architectures, with CNNs excelling at capturing local harmonic patterns and RNNs addressing sequential dependencies. Hybrid models~\cite{korzeniowski2016fully,lanz2021automatic} combining CNNs with Conditional Random Fields or RNNs demonstrated significant performance improvements by integrating learned auditory features with sequence modeling capabilities. Fully convolutional models and Bidirectional LSTMs~\cite{jiang2019large, wu2018automatic} further showcased the potential of deep learning to address ACR's complexities. To address the limitations of CNNs and RNNs, attention mechanisms and Transformer-based architectures emerged as key innovations~\cite{vaswani2017attention}. Transformers, leveraging self-attention mechanisms, dynamically weigh sequence elements, enabling robust modeling of global context and long-term harmonic dependencies~\cite{gulati2020conformer}. Recent advancements, such as the Bi-Directional Transformer for Chord Recognition~\cite{park2019bi}, demonstrated the effectiveness of these architectures in sequence segmentation and chord classification.

ACR systems must also address challenges posed by large vocabularies and class imbalance~\cite{pauwels2019acr}, particularly the underrepresentation of rare chord qualities. Structured chord representations have proven effective for these challenges. By encoding chords as combinations of musically meaningful components, such as root notes, bass positions, and chord qualities, these methods align with music theory while reducing learning complexity~\cite{mcfee2017structured}. For instance, the HPA system~\cite{ni2012end} models chords using latent variables, such as root-position form and bass note, within an HMM framework. Similarly, McFee and Bello ~\cite{mcfee2017structured} proposed a 36-dimensional binary vector encoding the root, bass, and chord degrees, facilitating a better modeling of chord relationships. Hierarchical and decompositional representations further improve model generalization across rare and extended chord types.

Recent innovations have introduced strategies to address class imbalance and improve model robustness. Jiang et al.~\cite{jiang2019large} proposed a reweighting loss strategy that assigns higher weights to underrepresented chord components, mitigating learning biases and ensuring a balanced recognition across chord classes. Semi-supervised methods and variational autoencoders~\cite{wu2020semi} have been used to effectively combine generative and discriminative techniques to maximize the use of unlabeled data. Curriculum learning~\cite{rowe2021curriculum} was also adopted to progressively introduce rare chords during training, leveraging hierarchical relationships between base and extended chord qualities to enhance classification performance.

Contrastive learning frameworks~\cite{li2024large}, combined with noisy-student models, have further enhanced the utilization of unlabeled data to improve recognition accuracy for both frequent and rare chord classes. These approaches employ contrastive learning for robust representation extraction and pseudo-labeling with data augmentation to expand training datasets and mitigate annotation biases. Evaluations on benchmark datasets, such as the Billboard and Humphrey-Bello collections~\cite{mcfee2017structured, humphrey2015four}, demonstrated the efficacy of these methods in addressing large vocabulary challenges while achieving balanced classification.

Building on these advancements, the proposed ChordFormer model fuses convolutional layers and self-attention mechanisms within a conformer-based architecture. This hybrid design captures both local and global dependencies in chord sequences. By incorporating a reweighted loss function and structured chord representations, the model addresses the dual challenges of class imbalance and large vocabulary complexity. Experimental results highlight the effectiveness of the proposed approach, showing a significant performance improvement in recognizing rare and extended chord types  with respect to state-of-the-art models. Furthermore, the integration of musical semantics into structured label representations proved to be effective in bridging the gap between theoretical music knowledge and practical applications.

\section{Methodology}\label{arc}

This section presents the proposed ChordFormer architecture for large-vocabulary chord recognition. The methodology integrates several components, including a structured representation of chord labels, pre-processing, feature extraction using conformer blocks, a decoding model based on Conditional Random Fields, and a class re-weighting mechanism to address imbalance issues. Each component is detailed in the following subsections.

\subsection {Chord Labels Structured Representation for Recognition}\label{csr}

An important insight from the literature~\cite{mcfee2017structured, ni2012end, jiang2019large} is that the effectiveness of models in chord classification tasks heavily depends on the chord representation employed in terms of embeddings. A good chord representation not only enhances classification performance but can also help in addressing the challenge of class imbalance in large chord vocabularies. Representing a chord by its meaningful components, such as root, triad, bass, and extensions (7\textsuperscript{th}, 9\textsuperscript{th}, 11\textsuperscript{th}, and 13\textsuperscript{th}), captures the hierarchical and semantic relationships among the notes that compose complex chords. It further accommodates all possible inversions through bass identification. Such a representation not only mitigates the complexity associated with large chord vocabularies, but also standardizes component labels for a precise and interpretable recognition.

The chord representation utilizes the following components and labels:
\begin{itemize}
    \item \textbf{Root + triad:} where root $\in$\{N, C, C\#/D$\flat$, ..., B\} and triad $\in$\{N, major, minor, sus4, sus2, diminished, augmented\}
    \item \textbf{Bass:} \{N, C, C\#/D$\flat$, ..., B\}
    \item \textbf{7\textsuperscript{th}:} \{N, 7, $\flat$7, $\flat\flat$7\}
    \item \textbf{9\textsuperscript{th}:} \{N, 9, \#9, $\flat$9\}
    \item \textbf{11\textsuperscript{th}:} \{N, 11, \#11\}
    \item \textbf{13\textsuperscript{th}:} \{N, 13, $\flat$13\}
\end{itemize}

In the notation presented above, it is worth observing that:
\begin{itemize}
\item  The ``N'' class included in each component denotes the absence of that component.
\item The double flat seventh ($\flat\flat$7) denotes the sixth and it is used to maintain a theoretical consistency with certain notations encountered in complex modulations.
\item The sixth and thirteenth extensions refer to the same absolute note, but their naming depends on the harmonic context and chord function: while a sixth chord refers to a major triad with an added sixth (e.g., C6 = C-E-G-A) without a seventh, a thirteenth chord normally includes the dominant seventh ($\flat$7) and the ninth ($9$), extending beyond the octave (e.g., C13 = C-E-G-B$\flat$-D-A).
\end{itemize}

The proposed ChordFormer represents chords using a six-dimensional vector, with each dimension corresponding to a specific chord component. In particular, the first dimension encodes the combination of the root and triad types (e.g., C\# augmented, D minor). 
The second to sixth dimensions represent the bass (capturing inversions), 7\textsuperscript{th}, 9\textsuperscript{th}, 11\textsuperscript{th}, and 13\textsuperscript{th} extensions, respectively. Each component label in the output is encoded using a one-hot representation, enabling the model to focus on specific chord components. This structured approach, proposed by Jiang et al~\cite{jiang2019large}, has been selected for improving the interpretability of the classification output and facilitating precise recognition of chords, even in scenarios involving rare extensions or complex inversions. This approach also allows a more consistent comparison of the results with respect to the current state-of-the-art methods.

\subsection {ChordFormer Architecture Overview}

\begin{figure*}[ht!]
    \centering
    \includegraphics[width=\linewidth]{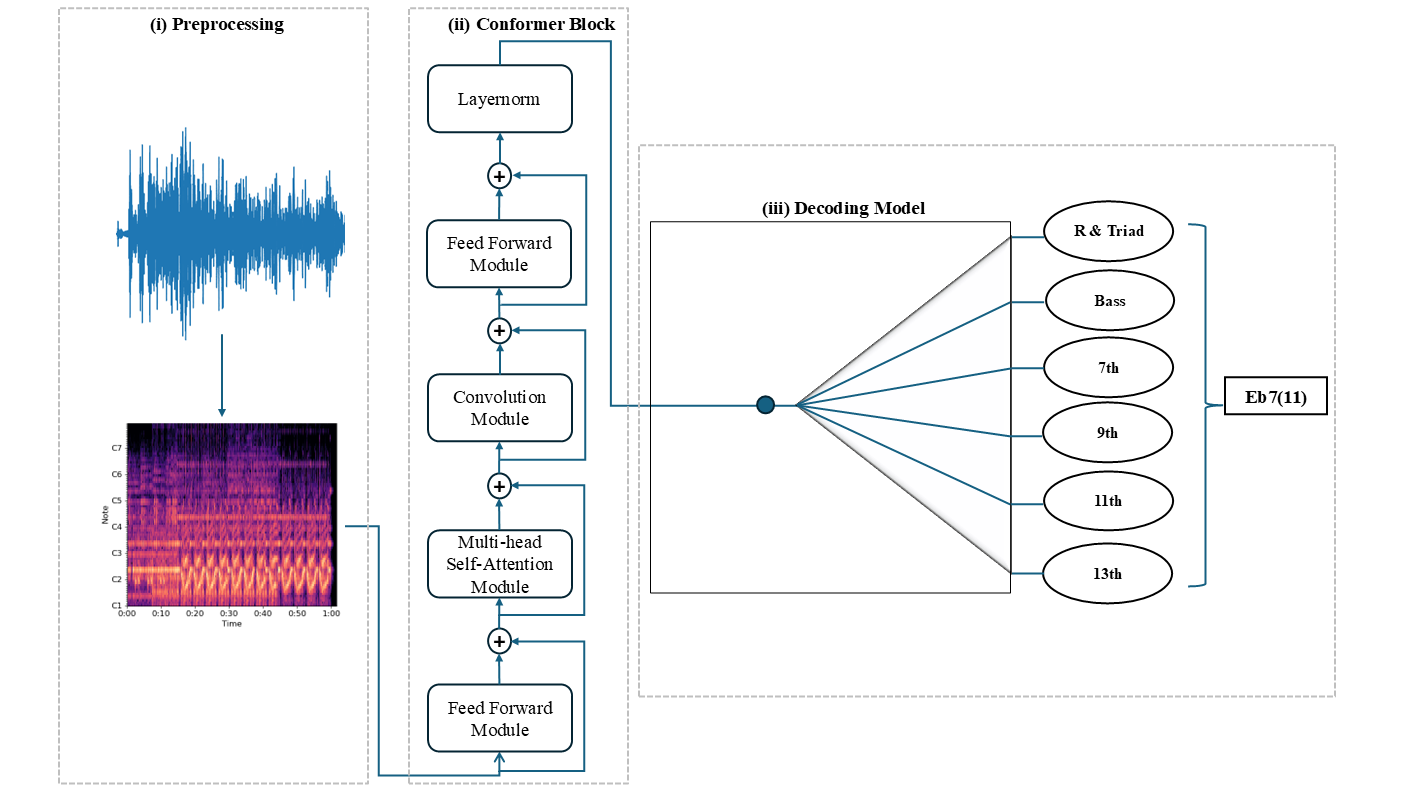}
    \caption{
    \textbf{Overview of the ChordFormer Architecture.} 
    The ChordFormer architecture consists of three primary components:
    (i) \textbf{Preprocessing Module:} Converts audio signals into Constant-Q Transform (CQT) representations to extract low-level time-frequency features essential for chord recognition.
    (ii) \textbf{Conformer Block:} Processes the CQT features to generate frame-wise activations using a combination of convolutional layers and self-attention mechanisms. This block leverages contextual frames to enhance accuracy in chord prediction.
    (iii) \textbf{Decoding Model:} Interprets the activations from the Conformer Block and generates the final chord sequence by decoding the structural chord components.
    }
    \label{fig:system}
\end{figure*}

The ChordFormer architecture integrates three main components: a pre-processing module, a conformer block, and a decoding model. They work together to achieve efficient chord recognition. The pre-processing module converts audio into a spectrogram computed by the Constant Q Transform (CQT)~\cite{Brown91}, which is more suitable for music, since it provides a time-frequency representation where frequency bins are logarithmically spaced, having higher resolution for lower frequencies and lower resolution for higher frequencies. The CQT spectrogram is then processed by the conformer block that produces frame-wise activations that capture chord component values. By leveraging contextual frames, the conformer block enhances the accuracy of the predictions. To address the context-dependent nature of chord recognition, the ChordFormer incorporates a modified Transformer design with conformer blocks~\cite{zhang2020transformer, karita2019comparative}, which enable advanced feature extraction. Finally, the decoding model interprets these activations to generate the final chord sequence. An overview of the ChordFormer architecture is illustrated in Figure~\ref{fig:system}.

\subsection {Feature extraction using conformer blocks}

The Conformer block, introduced by Gulati et al.~\cite{gulati2020conformer}, has become a state-of-the-art architecture for speech recognition. It combines Convolutional Neural Networks (CNNs) for capturing local patterns and self-attention mechanisms for long-range dependencies, making it highly effective for sequence modeling tasks. A Conformer block includes a feed-forward module, a self-attention module, a convolution module, and a final feed-forward module. Sections~\ref{multi}, \ref{conv}, and \ref{ff} introduce the self-attention, convolution, and feed-forward modules, respectively. Lastly, the process of combining these sub-blocks is explained in Section~\ref{conf}.
\subsubsection{Multi-Headed Self-Attention}\label{multi}
Accurate chord recognition requires integrating information from the target frame, its surrounding frames, and correlated frames. Traditional architectures, such as CNNs and Recurrent Neural Networks (RNNs), are effective at leveraging contextual information. However, self-attention mechanisms offer distinct advantages, making them particularly well-suited for this task.

One significant strength of self-attention is its ability to selectively focus on relevant frames, resulting in an adaptive receptive field. Unlike CNNs, which operate with a fixed kernel size, self-attention dynamically adjusts its focus based on the input. In contrast, a self-attention mechanism can dynamically attend to relevant frames, effectively isolating the target irrespective of its position.

An additional benefit of self-attention is its effectiveness in accurately encapsulating long-term dependencies. While RNNs can process distant information, they lack direct access to the related frames, which reduces their efficiency. CNNs partially address this issue by increasing the number of layers or kernel size, but these approaches come with inherent limitations: deeper CNNs lose weight-sharing efficiency, and RNNs often encounter information degradation over extended sequences. Self-attention resolves these challenges by enabling immediate access to any frame within the sequence, regardless of its distance from the target frame with an additional computational cost.

The proposed approach leverages Multi-Headed Self-Attention (MHSA), enhanced by the relative sinusoidal positional encoding method—a significant improvement introduced in Transformer-XL \cite{dai2019transformer}. This encoding scheme enhances the self-attention mechanism’s generalization capability by explicitly modeling relative spatial and temporal correlations. Consequently, the encoder can effectively contextualize temporal information, resulting in a more robust representation of the input sequence.

In MHSA, the input features for each time frame are divided into \( n_h \) segments, corresponding to individual attention heads. For an input matrix \( \mathbf{Z} \), the MHSA mechanism is defined as:

\begin{equation}
\text{Multihead} = \text{Concat}(h_1, \ldots, h_{n_h})W_O
\label{eq:multihead}
\end{equation}

\noindent
where $h_1$ represents the output of the first attention head, and $W_O$ is the weight matrix of the final fully connected layer that projects the concatenated outputs of all attention heads into the desired dimensional space.

The query, key, and value matrices for the \( j \)-th attention head are denoted as \( Q_j = (\mathbf{Z}W_Q)_j \), \( K_j = (\mathbf{Z}W_K)_j \), and \( V_j = (\mathbf{Z}W_V)_j \), respectively. These matrices are computed by projecting the input matrix \( \mathbf{Z} \) through fully connected layers \( W_Q \), \( W_K \), and \( W_V \), ensuring that each head processes a unique representation of the input features.

The attention function for each head \( h_j \) is defined as:

\begin{equation}
\text{Attention}(Q, K, V) = \text{softmax}\left(\frac{QK^\top}{\sqrt{d_K}}\right)V
\label{eq:attention}
\end{equation}

In this formulation, \( d_K \) represents the dimensionality of the key vectors. The scaling factor \( \sqrt{d_K} \) is introduced to mitigate instability during training by preventing excessively large values in the dot-product operation.

The softmax activation function normalizes the attention scores into probabilities, ensuring that each key-value pair contributes appropriately. After concatenating the outputs from all \( n_h \) attention heads, the result is processed through a fully connected layer \( W_O \), projecting the combined output (of size \( n_h \times d_{V_j} \)) into the desired dimension.

To further enhance training stability, pre-norm residual connections~\cite{wang2019learning, nguyen2019transformers} and dropout are incorporated. These layers help mitigate overfitting and improve gradient flow, particularly in deeper architectures. Figure~\ref{fig:block}(ii) illustrates the multi-head self-attention block’s structure, highlighting its modular design and data flow.

\subsubsection{Position-Wise Convolution}\label{conv}

To efficiently capture and utilize neighboring feature information within each time frame, a position-wise convolutional network is employed instead of the fully connected feedforward network present in the original Transformer architecture. Building on prior research~\cite{wu2020lite}, the convolutional module is designed to begin with a gating mechanism~\cite{dauphin2017language}, incorporating a pointwise convolution followed by a gated linear unit (GLU). This initialization is subsequently enhanced with a one-dimensional depth-wise convolutional layer, further refining the module’s capacity for feature extraction and contextual representation.

To ensure consistency in feature size and sequence length, the dimensions of both input and output channels are kept uniform across the convolutional block. By incorporating contextual information from adjacent frames, this position-wise convolutional structure effectively captures boundary transitions and smooth chord sequences. Batch normalization is applied immediately after the convolutional layers to enhance training stability. Figure~\ref{fig:block}(iii) presents a detailed scheme of the convolutional block, illustrating its structure and role within the overall framework.

\subsubsection{Feed-Forward Block}\label{ff}

The FeedForward Block, as introduced in the Transformer architecture~\cite{vaswani2017attention}, is positioned after the MHSA layer and serves a pivotal role in enhancing the model's expressive capacity. This module consists of two linear transformations separated by a nonlinear activation function, forming a robust framework widely employed in Transformer-based models, including those designed for automatic speech recognition~\cite{dong2018speech}. A residual connection is applied across the feedforward layers, accompanied by layer normalization, which ensures training stability and facilitates efficient convergence.

ChordFormer employs pre-norm residual units~\cite{wang2019learning, nguyen2019transformers}, where layer normalization is applied to both the residual unit and the input prior to the initial linear transformation. This adjustment facilitates improved gradient flow, particularly in deeper architectures, thereby enhancing overall training stability. Additionally, neurons use the Swish activation function~\cite{ramachandran2017searching}, which has demonstrated superior efficacy compared to conventional activation functions, alongside Dropout for regularization. These strategies collectively mitigate overfitting and provide the model with enhanced robustness.

\begin{figure*}[ht!]
    \centering
    \includegraphics[width=0.8\linewidth]{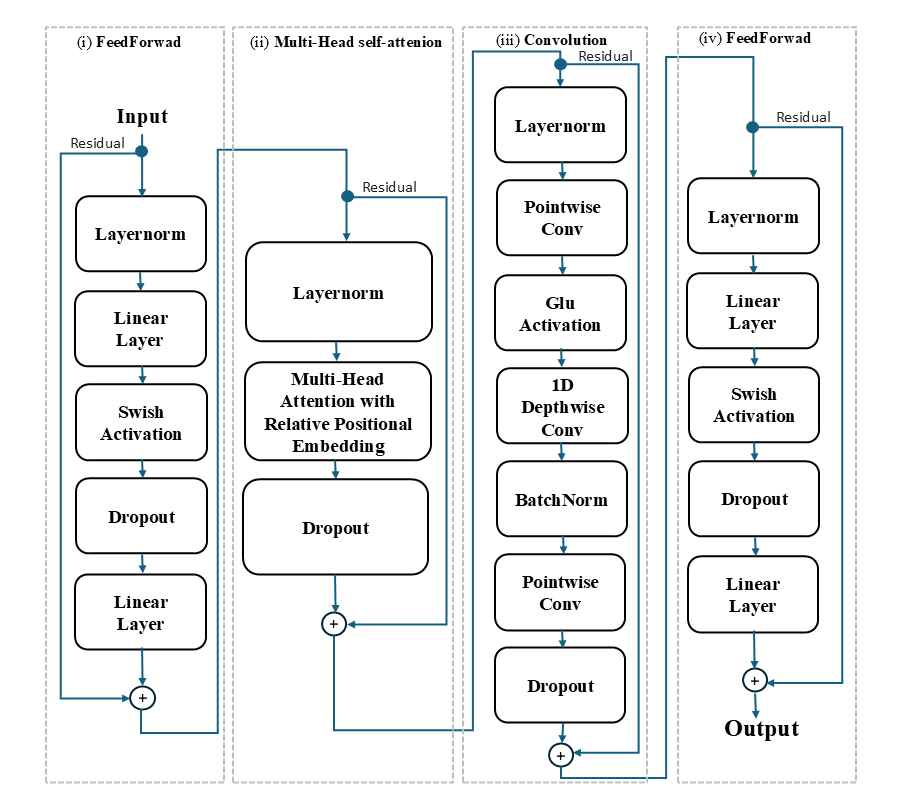}
    \caption{\textbf{Architecture Overview of Key Modules in ChordFormer Model.} 
    (i) \textbf{Feedforward Module}: Consists of a pre-normalization layer followed by a linear layer, Swish activation, and a dropout mechanism. Another linear layer projects back to the model dimensions. A residual connection enhances feature propagation. 
    (ii) \textbf{Multi-Head Self-Attention Module}: Incorporates multi-head self-attention with relative positional embeddings to capture global dependencies. Pre-normalization ensures stable gradient flow, and a residual connection is included to preserve input information. 
    (iii) \textbf{Convolution Module}: Features a pointwise convolution with an expansion factor of 2, coupled with a Gated Linear Unit (GLU) activation. This is followed by a 1-D depthwise convolution, batch normalization, and a Swish activation layer. Dropout is applied to ensure regularization, and a residual connection completes the module. 
    (iv) \textbf{Feedforward Module}: Reiterates the feedforward  structure with pre-normalization, Swish activation, and dropout for robust learning and dimensionality preservation.
    }
    \label{fig:block}
\end{figure*}

By integrating these techniques, the proposed method achieves greater stability during training and significantly enhances its generalization capabilities, making it particularly well-suited for complex sequence modeling tasks. Figure~\ref{fig:block}(i) illustrates the architecture of the Feed Forward Network (FFN) module and its integration within the Transformer framework.

\subsubsection{Conformer Block}\label{conf}

The Conformer block builds upon the previously described modules by incorporating a unique design that places the MHSA and the Convolution module between two Feed-Forward modules, as shown in Figure~\ref{fig:system}. This design extends the methodology introduced earlier by integrating components to balance global dependencies and local contextual information effectively. The detailed workflow of the Conformer block is illustrated in Figure~\ref{fig:block}. The architecture is inspired by Macaron-Net~\cite{lu2019understanding}, which replaces the conventional feedforward network in Transformer blocks with two half-step feedforward layers—one positioned before the attention layer and the other after it.

Following the principles of Macaron-Net, the proposed ChordFormer implementation employs residual weights that are half-step in nature within the FFN modules. Moreover, the second FFN module is followed by a final layer normalization phase to enhance both stability and performance. Given an input \( \mathbf{Z}_i \) to the Conformer block \( i \), the output \( \mathbf{Z}_i^{(o)} \) is calculated as follows:

\begin{equation}
\tilde{\mathbf{Z}}_i = \mathbf{Z}_i + \frac{1}{2} \text{FFN}(\mathbf{Z}_i)
\end{equation}

\begin{equation}
\mathbf{Z}_i^{(a)} = \tilde{\mathbf{Z}}_i + \text{MHSA}(\tilde{\mathbf{Z}}_i)
\end{equation}

\begin{equation}
\mathbf{Z}_i^{(c)} = \mathbf{Z}_i^{(a)} + \text{Conv}(\mathbf{Z}_i^{(a)})
\end{equation}

\begin{equation}
\mathbf{Z}_i^{(o)} = \text{LayerNorm}\left(\mathbf{Z}_i^{(c)} + \frac{1}{2} \text{FFN}(\mathbf{Z}_i^{(c)})\right)
\end{equation}

Here, \( \text{FFN} \) represents the Feed-Forward module, \( \text{MHSA} \) denotes the Multi-Head Self-Attention module and \( \text{Conv} \) corresponds to the Convolution module, which focuses on local feature extraction and pattern recognition, as detailed in earlier sections. Finally, \( \text{LayerNorm} \) refers to the layer normalization technique applied at the end to stabilize training and improve overall performance. This architecture strategically allocates the feedforward layers around the attention and convolutional modules, extending the capabilities described in the previous section. By doing so, it efficiently balances the model’s effectiveness to encapsulate global dependencies and local contextual information, making it particularly well-suited for music chord recognition.

\subsection {Loss Function}

The Conformer block outputs a hidden vector of $M_c$ components, one for each time frame $t$. To transform this hidden state into meaningful chord predictions, a linear unit is applied, projecting the hidden state into six separate vectors, denoted as $\mathbf{S}^{t,j}_m$, where $j = 1, 2, \dots, 6$. Each vector corresponds to a distinct component of the chord sequence, as described in section \ref{csr}, which includes components like root, triad, bass, and extensions (7\textsuperscript{th}, 9\textsuperscript{th}, 11\textsuperscript{th}, and 13\textsuperscript{th}).

A softmax function is then applied to these vectors to compute the activation probabilities for the corresponding chord components. The activation probability $\beta_m^{(t, j)}$ for the $j$-th component at time $t$ is computed as:

\begin{equation}
\beta_m^{(t, j)} = 
\frac{\exp\left(\mathbf{S}^{(t,j)}_m\right)}{\sum\limits_{m'=1}^{M_j} \exp\left(\mathbf{S}^{(t,j)}_{m'}\right)}
\quad 
\begin{aligned}
&\forall j = 1, \dots, 6, \\
&\forall m = 1, \dots, M_j
\end{aligned}
\end{equation}

\noindent
where $\beta_m^{(t, j)}$ represents the activation probability of class $m$ for the $j$-th chord component at frame $t$, $\mathbf{S}^{t,j}_m$ is the corresponding score for class $m$, and $M_j$ is the vocabulary size of the $j$-th chord component. This structured approach ensures that the predicted activations are normalized and interpretable, enabling effective modeling of complex harmonic relationships within chords.

To optimize the model, a weighted cross-entropy loss is employed to account for class imbalance. Let $\mathbf{Z} = \{\mathbf{Z}^{(1)}, \dots, \mathbf{Z}^{(T)}\}$ be the ground-truth chord sequence, where $\mathbf{Z}^{(t)}$ is the chord vector at time $t$, and let $z_j^{(t)}$ be the ground-truth class index for the $j$-th component at time $t$. Then, the loss function is defined as:

\begin{equation}
\mathcal{L} = - \sum_{t=1}^{T} \sum_{j=1}^{6} \sum_{m=1}^{M_j} w_m^{(j)} \, \mathbb{I}[m = z_j^{(t)}] \log \beta_m^{(t, j)}
\end{equation}

\noindent
where \( w_m^{(j)} \) is the weight assigned to class \( m \) of the \( j \)-th component to address class imbalance, \( \mathbb{I}[m = z_j^{(t)}] \) is the indicator function that equals 1 if the predicted class matches the ground-truth class and 0 otherwise, and \( \beta_m^{(t, j)} \) is the activation probability predicted for the respective class. This weighted loss ensures that rare chord classes are appropriately considered during training, enabling a balanced performance across all components.

\subsection {Addressing Data Imbalance through Class Re-weighting}

During the training phase, the issue of class imbalance became evident, even though the model does not directly classify individual chords. Certain chord extensions are poorly represented in the training set, resulting in an imbalance with an excess of majority class samples and a shortage of minority class samples. This imbalance skews the training process, causing the model to prioritize optimizing performance for classes with larger sample sizes while disadvantaging underrepresented classes.

An additional critical factor exacerbating bias in automatic chord recognition systems is the inherent ambiguity in audio chord annotation. Even among human experts, determining the "ground truth" chord label for a specific audio segment often involves subjective interpretation. This ambiguity can result in overlapping class boundaries, generating uncertainty and reducing classification accuracy. Such ambiguity disproportionately impacts classes with lower prior probabilities, further limiting the overall performance.

To overcome these limitations, we applied a class re-weighting strategy that assigns a class-specific weight factor \( w_m^{(j)} \) for each possible value of a chord component, using approach presented by Jiang et al.~\cite{jiang2019large}. This re-weighting approach mitigates class imbalance by amplifying the contribution of underrepresented classes in the loss function during the training phase. The weight factor is defined as follows:

\begin{equation}
w_m^{(j)} = \min \left\{ \left( \frac{n_m^{(j)}}{\max_{m'} n_{m'}^{(j)}} \right)^{-\gamma}, w_{\text{max}} \right\},
\end{equation}

\noindent
where \( n_m^{(j)} \) represents the total number of training samples associated with class \( m \) in the \( j \)-th component. The balancing factor \( \gamma \), constrained to the range \( 0 \leq \gamma \leq 1 \), regulates the degree of weighting adjustment, with smaller values resulting in more equitable weight distribution across classes. Additionally, \( w_{\text{max}} \) serves as the clamping parameter, set to \( w_{\text{max}} \geq 1 \), which prevents the weight values from reaching excessively large magnitudes.

By adopting this weighting strategy, underrepresented classes with fewer samples are assigned larger weights, effectively mitigating bias in the training process. Lower values of \( \gamma \) yield a more uniform weight distribution, whereas higher values allow the weights to more closely reflect the sample distribution. This approach ensures that infrequent chord classes are adequately emphasized during training, thereby reducing the adverse effects of data imbalance and annotation ambiguity. The proposed ChordFormer leverages this re-weighting strategy to establish a more balanced training procedure, enhancing its robustness and accuracy across a diverse range of chord classes.
\subsection {Decoding Strategy for Chord Recognition}

A straightforward approach to decoding the final chord sequence from the activation probabilities involves selecting the class that has the highest probability for each component at every frame. However, this method often results in excessive transitions between adjacent chords, as it lacks a mechanism to impose penalties for immediate changes. Additionally, this approach provides limited control over the output chord vocabulary, which can lead to inconsistencies in the predictions. To overcome these challenges, ChordFormer incorporates a linear Conditional Random Field (CRF) for decoding the final chord sequence. This technique effectively balances activation probabilities with temporal smoothness, ensuring more coherent chord transitions. The CRF models the probability of a chord sequence \( \mathbf{Z} \) given the audio features \( \mathbf{X} \) using the following formulation:

\begin{equation}
P(\mathbf{Z} \mid \mathbf{X}) \propto \phi(\mathbf{Z}^{(1)}, \mathbf{X}) \prod_{t=2}^{T} \phi(\mathbf{Z}^{(t)}, \mathbf{X}) \psi(\mathbf{Z}^{(t-1)}, \mathbf{Z}^{(t)}).
\end{equation}

Here, \( \phi \) is the observation potential function that evaluates the compatibility between the audio features and the chord predictions, while \( \psi \) is the transition potential function that enforces temporal coherence by penalizing immediate transitions.

The observation potential function is defined as:

\begin{equation}
\phi(\mathbf{Z}^{(t)}, \mathbf{X}) = \exp\left( \sum_{j=1}^{6} \sum_{m=1}^{M_j} \mathbb{I}[m = z_j^{(t)}] \log \beta_m^{(t, j)} \right).
\end{equation}

The transition potential function that regulates the smoothness of the chord sequence is expressed as:

\begin{equation}
\psi(\mathbf{Z}^{(t-1)}, \mathbf{Z}^{(t)}) = \exp\left( -\gamma \cdot \mathbb{I}[\mathbf{Z}^{(t-1)} \neq \mathbf{Z}^{(t)}] \right),
\end{equation}

\noindent
where \( \gamma \) is a hyperparameter that controls the penalty for transitions between different chords. Higher values of \( \gamma \) enforce smoother transitions, reducing immediate changes in the chord sequence.

By integrating these potentials, the CRF-based decoding method provides a robust balance between prediction accuracy and temporal coherence. This method not only reduces excessive chord transitions but also allows explicit control over the output chord vocabulary, making it particularly well-suited for large-vocabulary chord recognition tasks~\cite{wu2018automatic}.

\section{Experiments}\label{exp}

A set of experiments have been conducted to evaluate the performance of the proposed ChordFormer model in comparison to other baseline models from the literature. This section explains
how data are organized and preprocessed, describes the model's training setup and the evaluation metrics for assessing models' performance, and presents the achieved results, highlighting the differences with respect to the other models.

\subsection {Dataset Description and Preprocessing}

The evaluation of the proposed model has been carried out using the dataset by Humphrey and Bello~\cite{mcfee2017structured, humphrey2015four} comprising 1,217 songs obtained from the Isophonics, Billboard, and MARL collections. To ensure a rigorous and unbiased evaluation, we implemented a 5-fold cross-validation strategy, using the same training, validation, and testing ratio defined in~\cite{humphrey2015four}. Specifically, 60\% of the dataset has been reserved for training, 20\% for validation, and 20\% for testing. Such a partitioning is consistent with prior studies, thus enabling a direct and reliable comparison of results while maintaining methodological integrity.

The CQT spectrograms were generated from the audio data using the Librosa library, configured with a sampling rate of 22,050 Hz and a hop length of 512. The frequency range extends from note C1 (inclusive) to C8 (exclusive), utilizing 36 bins per octave, resulting in a total of 252 CQT bins. After that, the spectrogram was transformed into the decibel scale using Librosa’s \texttt{amplitude\_to\_db} function, with normalization relative to the maximum amplitude. This step enhances interpretability by representing the data on a logarithmic scale, which aligns with the human auditory perception of sound intensity. To enhance the model's robustness, data augmentation was applied to the training set using pitch-shifting techniques. These pitch shifts range from \(-5\) to \(+6\) semitones, generating augmented features through direct modifications of the CQT spectrograms. The corresponding annotated chord labels were adjusted to align with the altered audio attributes. This augmentation strategy ensures a more diverse and representative training dataset while preserving precise chord annotations. By incorporating this approach, the model's ability to generalize effectively across varied musical contexts was significantly improved, advancing its performance in music chord recognition tasks.

\subsection {Training ChordFormer and Hyperparameters}

Table~\ref{tab:chordconformer} summarizes the hyperparameters employed for the Conformer block, following a 5-fold cross-validation process. These hyperparameters were carefully tuned to achieve the best results, ensuring both robustness and accuracy in the model's performance. The AdamW optimizer was used to train the neural network, incorporating a dynamic learning rate schedule to enhance optimization. The initial learning rate was set at \( 1 \times 10^{-3} \) and reduced by 90\% if no improvement was observed in the validation loss over five consecutive epochs. Training concludes when the learning rate drops below \( 1 \times 10^{-6} \), ensuring efficient computation and minimizing overfitting risks.

Every epoch involves the random extraction of a 1,000-frame segment (approximately 23.2 seconds) from every song to introduce temporal diversity. Mini-batches were formed by combining 24 such segments, effectively balancing computational efficiency with data variability to facilitate robust model training.

\begin{table}[ht!]
\centering
\caption{ChordFormer Parameters}
\label{tab:chordconformer}
\begin{tabular}{@{}ll@{}}
\toprule
\textbf{Parameter}                & \textbf{Value} \\ \midrule
\texttt{input\_dim}               & 256            \\
\texttt{num\_heads}               & 4              \\
\texttt{ffn\_dim}                 & 1024           \\
\texttt{num\_layers}              & 4              \\
\texttt{depthwise\_conv\_kernel\_size} & 31             \\
\texttt{output\_dim}              & 100            \\ \bottomrule
\end{tabular}
\end{table}


\begin{table*}[!ht]
\centering
\renewcommand{\arraystretch}{1.3} 
\caption{Comparison of Weighted Chord Symbol Recall (WCSR) Across Different Baseline Models on Various Metrics.}
\label{tab:model_comparison}
\begin{tabular}{lccccccc}
    \toprule
    \textbf{Model} & \textbf{Root} & \textbf{Thirds} & \textbf{MajMinor} & \textbf{Triads} & \textbf{Sevenths} & \textbf{Tetrads} & \textbf{MIREX} \\
    \midrule
    BTC+CNN \cite{rowe2021curriculum}        & 54.28   & 47.94   & 49.00    & 44.67    & 37.99    & 34.01   & 47.94    \\
    Transformer \cite{vaswani2017attention}   & 78.55    & 72.91    & 74.24    & 67.75    & 57.42    & 51.46    & 72.23    \\
    CNN \cite{korzeniowski2016fully}               & 81.76    & 78.69    & 80.88    & 74.13    & 67.27    & 60.48    & 79.42    \\
    Transformer+CNN    & 82.40    & 79.40    & 81.67    & 74.88    & 67.84    & 61.04    & 80.22    \\
    CNN+BLSTM \cite{jiang2019large} & 83.39 & 80.04 & 82.62 & 75.91 & 69.78 & 62.87 & 81.52 \\
    \textbf{ChordFormer-R}         & \textbf{83.87} & \textbf{80.54} & \textbf{81.86} & \textbf{76.02} & \textbf{69.80} & \textbf{63.48} & \textbf{82.98} \\
    \textbf{ChordFormer}         & \textbf{84.69} & \textbf{81.75} & \textbf{84.09} & \textbf{77.55} & \textbf{72.28} & \textbf{65.32} & \textbf{83.62} \\
    \bottomrule
\end{tabular}
\end{table*}



\begin{table*}[!ht]
\centering
\renewcommand{\arraystretch}{1.3} 
\caption{Comparison of \( \text{acc}_{\text{frame}} \) and \( \text{acc}_{\text{class}} \) between CNN+BLSTM and ChordFormer Models}
\label{tab:accuracy_comparison}
\begin{tabular}{lcccc}
    \toprule
    \textbf{Re-weighting Value} & 
    \textbf{acc\(_{\text{frame}}\)} & \textbf{acc\(_{\text{frame}}\)} & 
    \textbf{acc\(_{\text{class}}\)} & \textbf{acc\(_{\text{class}}\)} \\
    & \textbf{CNN+BLSTM \cite{jiang2019large}} & \textbf{\textbf{ChordFormer}} & \textbf{CNN+BLSTM \cite{jiang2019large}} & \textbf{\textbf{ChordFormer}} \\
    \midrule
    No Re-weighting & 0.7676 & \textbf{0.7877} & 0.3315 & \textbf{0.3884} \\
    (0.3, 10.0) & 0.7659 & \textbf{0.7801} & 0.3692 & \textbf{0.4426} \\
    (0.5, 10.0) & 0.7402 & \textbf{0.7772} & 0.3821 & \textbf{0.4406} \\
    (0.7, 20.0) & 0.7117 & \textbf{0.7416} & 0.3823 & \textbf{0.4471} \\
    (1.0, 20.0) & 0.6512 & \textbf{0.6994} & 0.3546 & \textbf{0.4157} \\
    \bottomrule
\end{tabular}
\end{table*}


\subsection {Evaluation metrics}

To facilitate a thorough comparison with existing models, the proposed ChordFormer was evaluated using the widely adopted \texttt{mir\_eval} library~\cite{raffel2014mir_eval}, a standard tool in automatic chord recognition research. The assessment includes Root Scores, Maj-Min Scores, Seventh Scores, Thirds Scores, Triad Scores, Tetrads Scores, and MIREX Scores. These metrics collectively provide a comprehensive evaluation of chord recognition performance, capturing various aspects of chord prediction. 

A key method used to measure these evaluation metrics is the Weighted Chord Symbol Recall (WCSR), proposed by Pauwels and Peeters~\cite{pauwels2013evaluating}, which quantifies the average frame-level accuracy of chord recognition. It is defined as

\begin{equation}
\text{WCSR} = \left(\frac{\sum_{i=1}^N z_i}{\sum_{i=1}^N Z_i}\right) \times 100
\label{eq:frame_accuracy}
\end{equation}

\noindent
where $N$ is the total number of songs in the evaluation dataset, $Z_i$ denotes the total duration of the $i$-th song, and $z_i$ corresponds to the duration of correctly predicted chord frames for the $i$-th song. 
This metric enables a quantitative understanding of how well a model performs across different musical contexts, providing a robust measure of its chord recognition capabilities.

\subsection {Evaluation metrics for large vocabulary}

To evaluate the model on a large vocabulary and address the challenges posed by imbalanced class, we followed the same approach proposed by Jiang et al.~\cite{jiang2019large} and tested it on a vocabulary $V$ consisting of 301 distinct chords, including Basic triads, Inverted triads, Seventh chords, Extended chords, Suspended chords, Slash chords, and a Non-chord class ($N$).
In particular, the evaluation incorporates two critical metrics: mean frame-wise accuracy (\(acc_{frame}\)) and mean class-wise accuracy (\(acc_{class}\)). These metrics are designed to measure both the overall performance and the fairness of chord recognition across different chord classes.

The mean frame-wise accuracy (\(acc_{frame}\)) is defined as:

\begin{equation}
acc_{frame} = \frac{\sum_{i=1}^n z_i}{\sum_{i=1}^n Z_i},
\end{equation}

\noindent
where \(n\) represents the total number of tracks in the test set, \(Z_i\) is the total number of frames in the \(i\)-th track, and \(z_i\) is the number of correctly predicted frames in the \(i\)-th track over the chord vocabulary \(V\).

The mean class-wise accuracy (\(acc_{class}\)) is defined as:

\begin{equation}
acc_{class} = \frac{1}{|V|} \sum_{v \in V} \frac{\sum_{i=1}^n z_i^v}{\sum_{i=1}^n Z_i^v},
\end{equation}

\noindent
where \(Z_i^v\) is the number of frames in the \(i\)-th track labeled with the ground-truth chord \(v\), and \(z_i^v\) is the number of frames correctly predicted as \(v\) in the \(i\)-th track. Here, \(|V|\) denotes the total number of unique chord classes in the vocabulary.


\begin{figure*}[ht!]
    \centering
    \includegraphics[width=\textwidth]{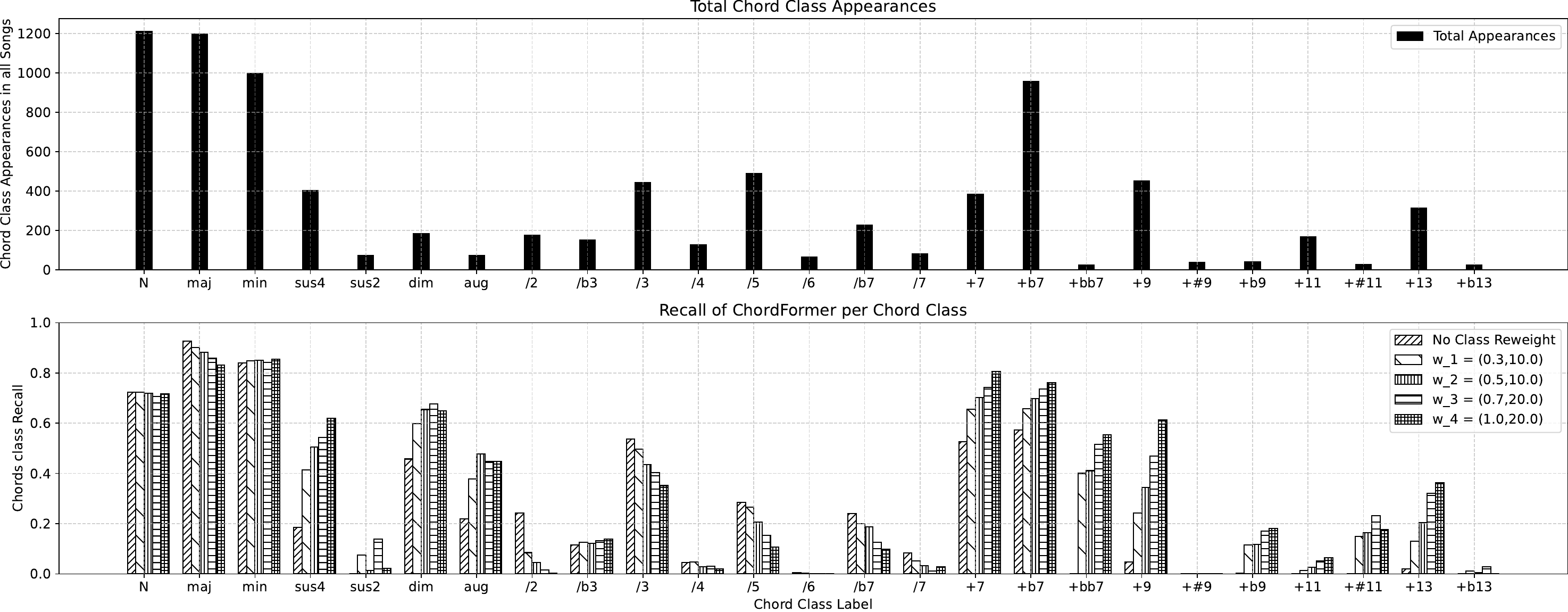} 
    \caption{Top: chord class appearance in the song dataset; Bottom: ChordFormer recall values for each chord class under varying re-weight factors (\( \gamma, w_{\text{max}} \)).}
    \label{fig:component-recall}
\end{figure*}


\section{Results and Discussion}\label{res}

To ensure a fair and consistent comparison, the proposed ChordFormer model and its variant ChordFormer-R (with reweighted loss) were evaluated against several baseline models using the same structural chord embeddings and training hyperparameters. The baseline models include
\begin{itemize}
  \item CNN: a convolutional neural network designed for automatic chord recognition proposed by Korzeniowski and Widmer~\cite{korzeniowski2016fully}.
  
  \item CNN+BLSTM: a convolutional neural network that incorporates a bidirectional long short-term memory (BLSTM) layer to enhance sequence modeling, as proposed by Jiang et al.~\cite{jiang2019large}.

  \item BTC+CNN: a model based on Bidirectional Self-Attention, evaluated by removing the position embedding layer and integrating self-attention with CNN, as proposed by Rowe and Tzanetakis~\cite{rowe2021curriculum}.

  \item Transformer: the original transformer model proposed by Vaswani~\cite{vaswani2017attention}, along with its combination with a CNN as a feature extractor (Transformer+CNN).
\end{itemize}

The WCSR results are reported in Table~\ref{tab:model_comparison} and clearly demonstrate the superior performance of ChordFormer and ChordFormer-R. In fact, ChordFormer achieves a Root accuracy of 84.69\%, MajMinor accuracy of 84.09\%, and a MIREX score of 83.62\%. These results highlight the Conformer block's ability to combine convolutional layers and self-attention mechanisms, enabling superior local and global feature modeling. Additionally, ChordFormer-R, which incorporates a reweighted loss function to address class imbalance, shows further improvements, particularly for rare chord classes. By assigning higher weights to underrepresented components during training, ChordFormer-R ensures a more balanced optimization process, showing its robustness in handling large-vocabulary chord recognition tasks effectively.

Further performance results are provided in Table~\ref{tab:accuracy_comparison}, which reports frame-wise accuracy (\( acc_{\text{frame}} \)) and class-wise accuracy (\( acc_{\text{class}} \)) across different re-weighting configurations for CNN+BLSTM and ChordFormer models. Without re-weighting, ChordFormer consistently outperforms CNN+BLSTM, achieving a higher \( acc_{\text{frame}} \) of 0.7877 compared to CNN+BLSTM’s 0.7676 and demonstrating a substantial improvement in \( acc_{\text{class}} \) (0.3884 vs. 0.3315). These results suggest that ChordFormer’s hybrid Conformer architecture excels in harmonic modeling, effectively capturing both local and long-range dependencies within musical sequences.

As class re-weighting increases, both models exhibit a decline in \( acc_{\text{frame}} \) due to the prioritization of rare chord classes. This decline is more pronounced for CNN+BLSTM, dropping from 0.7676 to 0.6512, while ChordFormer demonstrates greater resilience, maintaining \( acc_{\text{frame}} \) at 0.6994 even at higher re-weighting values (1.0, 20.0). Conversely, \( acc_{\text{class}} \) improves across all re-weighting configurations, with ChordFormer achieving the highest \( acc_{\text{class}} \) of 0.4471 at (0.7, 20.0), compared to CNN+BLSTM’s 0.3823. This highlights ChordFormer’s superior ability to generalize across rare chord classes, a critical advantage for real-world music information retrieval (MIR) applications.

\begin{figure}[t]
    \centering
    \includegraphics[width=1\linewidth]{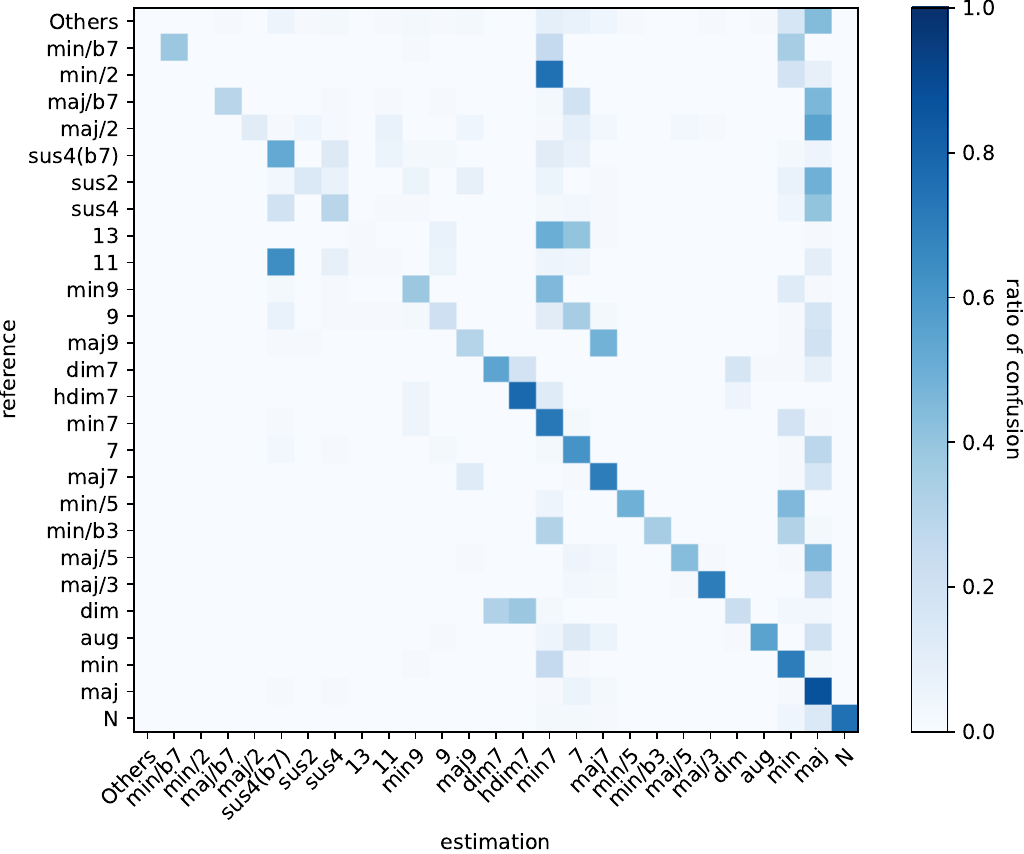}
    \caption{Confusion matrix of ChordFormer with re-weighting factors (\( \gamma = 0.5, w_{\text{max}} = 10 \))}
    \label{fig:enter-label}
\end{figure}

The confusion matrix reported in Figure~\ref{fig:enter-label} provides a detailed visualization of ChordFormer’s performance under re-weighting factors (\( \gamma = 0.5, w_{\text{max}} = 10 \)). Strong diagonal dominance reflects ChordFormer’s high accuracy in predicting common chord classes such as \textit{maj}, \textit{min}, and \textit{7}, which are frequent in Western music. However, moderate off-diagonal intensities indicate confusion among harmonically similar chords, such as \textit{min7}, \textit{hdim7}, and \textit{dim7}, or between extensions like \textit{9} and \textit{11}. While re-weighting strategies improve recognition for rare classes like \textit{maj9}, \textit{13}, and \textit{sus4(b7)}, they introduce slight trade-offs in overall frame-wise accuracy due to prioritization of minority classes.

To further evaluate the system’s performance on a large chord vocabulary, a comprehensive chord vocabulary encompassing all chord qualities in the dataset is adopted. The class-wise accuracy of each chord component is assessed during the decoding process, offering a detailed view of ChordFormer’s performance across a diverse range of harmonic structures. The graph in Figure~\ref{fig:component-recall} provides an in-depth analysis of chord component recall under various re-weighting configurations (\( \gamma, w_{\text{max}} \)). Common chord components such as \textit{maj} and \textit{min} consistently achieve high recall values, indicating ChordFormer’s strong ability to model frequent harmonic patterns. Rare components like \textit{dim}, \textit{aug}, and extended chords such as \textit{+11} and \textit{+13} show significant improvements under re-weighting strategies, with configurations like (\( 0.7, 20.0 \)) delivering noticeable gains.

However, trade-offs emerge, as higher re-weighting values (\( 1.0, 20.0 \)) lead to slight reductions in recall for dominant chord components, highlighting the challenge of balancing overall accuracy with enhanced recognition of rare chords. Configurations like (\( 0.5, 10.0 \)) and (\( 0.7, 20.0 \)) emerge as optimal trade-offs, enhancing recall for rare chords while maintaining strong performance for frequent ones. These evaluations underscore ChordFormer’s robustness in managing class imbalance and handling large-vocabulary chord recognition tasks, emphasizing the importance of parameter tuning to achieve balanced performance across diverse chord classes.

\section{Conclusions}\label{conc}

This work introduced ChordFormer, a novel Conformer-based architecture for large-vocabulary automatic chord recognition. By integrating convolutional layers with self-attention mechanisms, ChordFormer effectively captures both local spectral features and long-range harmonic dependencies, addressing key challenges in music information retrieval. The proposed model significantly outperforms state-of-the-art methods, demonstrating superior performance in both frame-wise accuracy and class-wise accuracy, particularly in recognizing rare and complex chord types.

To mitigate the long-tail distribution of chord datasets, a re-weighted loss function was employed, ensuring balanced learning across frequent and underrepresented chord classes. The results confirm that this strategy enhances the model’s generalizability without severely compromising frame-wise accuracy.

Furthermore, the use of structured chord representation aligns model’s predictions with music theory principles, thereby improving interpretability and robustness. Extensive evaluations on a large-scale chord recognition dataset illustrate that ChordFormer achieves state-of-the-art performance, surpassing transformer-based and recurrent models in handling intricate harmonic structures. The analysis of confusion matrices further highlights ChordFormer’s ability to distinguish harmonically similar chords while maintaining high accuracy in common chord classes.

While the results demonstrate ChordFormer’s potential in advancing automatic chord recognition, future research will explore adaptive re-weighting techniques to further optimize trade-offs between frame-wise and class-wise accuracy. Additionally, integrating self-supervised learning could enhance chord recognition by leveraging unlabeled music data. The findings of this study contribute to the ongoing development of deep learning techniques in music information retrieval, paving the way for more robust and interpretable chord recognition models in real-world applications.

\section*{Acknowledgment}

This work is partially supported by the Italian Ministry of University Research (MUR) in the context of the doctoral scholarship for cultural heritage, financed by the European Union - Next Generation EU and implemented with the Decree of the MUR n. 351 of 09 April 2022 and subsequent amendments.

\ifCLASSOPTIONcaptionsoff
  \newpage
\fi

\bibliographystyle{IEEEtran}
\bibliography{MTT_reveyrand}

\begin{thebibliography}{10}
\providecommand{\url}[1]{#1}
\csname url@rmstyle\endcsname
\providecommand{\newblock}{\relax}
\providecommand{\bibinfo}[2]{#2}
\providecommand\BIBentrySTDinterwordspacing{\spaceskip=0pt\relax}
\providecommand\BIBentryALTinterwordstretchfactor{4}
\providecommand\BIBentryALTinterwordspacing{\spaceskip=\fontdimen2\font plus
\BIBentryALTinterwordstretchfactor\fontdimen3\font minus \fontdimen4\font\relax}
\providecommand\BIBforeignlanguage[2]{{%
\expandafter\ifx\csname l@#1\endcsname\relax
\typeout{** WARNING: IEEEtran.bst: No hyphenation pattern has been}%
\typeout{** loaded for the language `#1'. Using the pattern for}%
\typeout{** the default language instead.}%
\else
\language=\csname l@#1\endcsname
\fi
#2}}
\renewcommand\BIBentryALTinterwordstretchfactor{4}

\bibitem{duran2020transcribing}
G.~Dur{\'a}n and P.~de~la Cuadra, ``Transcribing lead sheet-like chord progressions of jazz recordings,'' \emph{Computer Music Journal}, vol.~44, no.~4, pp. 26--42, 2020.

\bibitem{de2022measuring}
J.~de~Berardinis, A.~Cangelosi, and E.~Coutinho, ``Measuring the structural complexity of music: from structural segmentations to the automatic evaluation of models for music generation,'' \emph{IEEE/ACM transactions on audio, speech, and language processing}, vol.~30, pp. 1963--1976, 2022.

\bibitem{weiss2020local}
C.~Wei{\ss}, H.~Schreiber, and M.~M{\"u}ller, ``Local key estimation in music recordings: A case study across songs, versions, and annotators,'' \emph{IEEE/ACM Transactions on Audio, Speech, and Language Processing}, vol.~28, pp. 2919--2932, 2020.

\bibitem{du2023bytecover3}
X.~Du, Z.~Wang, X.~Liang, H.~Liang, B.~Zhu, and Z.~Ma, ``Bytecover3: Accurate cover song identification on short queries,'' in \emph{ICASSP 2023-2023 IEEE International Conference on Acoustics, Speech and Signal Processing (ICASSP)}.\hskip 1em plus 0.5em minus 0.4em\relax IEEE, 2023, pp. 1--5.

\bibitem{pauwels2019acr}
\BIBentryALTinterwordspacing
J.~Pauwels, K.~O’Hanlon, E.~Gómez, and M.~B. Sandler, ``20 years of automatic chord recognition from audio,'' in \emph{Proceedings of the 20th International Society for Music Information Retrieval Conference (ISMIR)}.\hskip 1em plus 0.5em minus 0.4em\relax Delft, Netherlands: International Society for Music Information Retrieval, November 4--8 2019, pp. 54--63. [Online]. Available: \url{https://zenodo.org/record/3527874}
\BIBentrySTDinterwordspacing

\bibitem{jiang2019large}
J.~Jiang, K.~Chen, W.~Li, and G.~Xia, ``Large-vocabulary chord transcription via chord structure decomposition.'' in \emph{ISMIR}, 2019, pp. 644--651.

\bibitem{rowe2021curriculum}
L.~O. Rowe and G.~Tzanetakis, ``Curriculum learning for imbalanced classification in large vocabulary automatic chord recognition.'' in \emph{ISMIR}, 2021, pp. 586--593.

\bibitem{de2023choco}
J.~de~Berardinis, A.~Mero{\~n}o-Pe{\~n}uela, A.~Poltronieri, and V.~Presutti, ``Choco: a chord corpus and a data transformation workflow for musical harmony knowledge graphs,'' \emph{Scientific Data}, vol.~10, no.~1, p. 641, 2023.

\bibitem{sheh2003chord}
A.~Sheh and D.~P. Ellis, ``Chord segmentation and recognition using em-trained hidden markov models,'' in \emph{International Symposium on Music Information Retrieval}, 2003.

\bibitem{ueda2010hmm}
Y.~Ueda, Y.~Uchiyama, T.~Nishimoto, N.~Ono, and S.~Sagayama, ``Hmm-based approach for automatic chord detection using refined acoustic features,'' in \emph{2010 IEEE International Conference on Acoustics, Speech and Signal Processing}.\hskip 1em plus 0.5em minus 0.4em\relax IEEE, 2010, pp. 5518--5521.

\bibitem{khadkevich2009use}
M.~Khadkevich and M.~Omologo, ``Use of hidden markov models and factored language models for automatic chord recognition.'' in \emph{ISMIR}, 2009, pp. 561--566.

\bibitem{korzeniowski2016feature}
F.~Korzeniowski and G.~Widmer, ``Feature learning for chord recognition: The deep chroma extractor,'' \emph{arXiv preprint arXiv:1612.05065}, 2016.

\bibitem{sigtia2015audio}
S.~Sigtia, N.~Boulanger-Lewandowski, and S.~Dixon, ``Audio chord recognition with a hybrid recurrent neural network.'' in \emph{ISMIR}, 2015, pp. 127--133.

\bibitem{zhou2015chord}
X.~Zhou and A.~Lerch, ``Chord detection using deep learning,'' in \emph{Proceedings of the 16th ISMIR Conference}, vol.~53, 2015, p. 152.

\bibitem{humphrey2012rethinking}
E.~J. Humphrey and J.~P. Bello, ``Rethinking automatic chord recognition with convolutional neural networks,'' in \emph{2012 11th International Conference on Machine Learning and Applications}, vol.~2.\hskip 1em plus 0.5em minus 0.4em\relax IEEE, 2012, pp. 357--362.

\bibitem{korzeniowski2016fully}
F.~Korzeniowski and G.~Widmer, ``A fully convolutional deep auditory model for musical chord recognition,'' in \emph{2016 IEEE 26th International Workshop on Machine Learning for Signal Processing (MLSP)}.\hskip 1em plus 0.5em minus 0.4em\relax IEEE, 2016, pp. 1--6.

\bibitem{mcfee2017structured}
B.~McFee and J.~P. Bello, ``Structured training for large-vocabulary chord recognition.'' in \emph{ISMIR}, 2017, pp. 188--194.

\bibitem{wu2018automatic}
Y.~Wu and W.~Li, ``Automatic audio chord recognition with midi-trained deep feature and blstm-crf sequence decoding model,'' \emph{IEEE/ACM Transactions on Audio, Speech, and Language Processing}, vol.~27, no.~2, pp. 355--366, 2018.

\bibitem{boulanger2013audio}
N.~Boulanger-Lewandowski, Y.~Bengio, and P.~Vincent, ``Audio chord recognition with recurrent neural networks.'' in \emph{ISMIR}.\hskip 1em plus 0.5em minus 0.4em\relax Curitiba, 2013, pp. 335--340.

\bibitem{deng2016hybrid}
J.-q. Deng and Y.-K. Kwok, ``A hybrid gaussian-hmm-deep learning approach for automatic chord estimation with very large vocabulary.'' in \emph{ISMIR}, 2016, pp. 812--818.

\bibitem{deng2017large}
------, ``Large vocabulary automatic chord estimation with an even chance training scheme.'' in \emph{ISMIR}, 2017, pp. 531--536.

\bibitem{vaswani2017attention}
A.~Vaswani, ``Attention is all you need,'' \emph{Advances in Neural Information Processing Systems}, 2017.

\bibitem{islam2024comprehensive}
S.~Islam, H.~Elmekki, A.~Elsebai, J.~Bentahar, N.~Drawel, G.~Rjoub, and W.~Pedrycz, ``A comprehensive survey on applications of transformers for deep learning tasks,'' \emph{Expert Systems with Applications}, vol. 241, p. 122666, 2024.

\bibitem{gulati2020conformer}
A.~Gulati, J.~Qin, C.-C. Chiu, N.~Parmar, Y.~Zhang, J.~Yu, W.~Han, S.~Wang, Z.~Zhang, Y.~Wu, \emph{et~al.}, ``Conformer: Convolution-augmented transformer for speech recognition,'' \emph{arXiv preprint arXiv:2005.08100}, 2020.

\bibitem{civit2022systematic}
M.~Civit, J.~Civit-Masot, F.~Cuadrado, and M.~J. Escalona, ``A systematic review of artificial intelligence-based music generation: Scope, applications, and future trends,'' \emph{Expert Systems with Applications}, vol. 209, p. 118190, 2022.

\bibitem{humphrey2015four}
E.~J. Humphrey and J.~P. Bello, ``Four timely insights on automatic chord estimation.'' in \emph{ISMIR}, vol.~10, 2015, pp. 673--679.

\bibitem{takuya1999realtime}
F.~Takuya, ``Realtime chord recognition of musical sound: Asystem using common lisp music,'' in \emph{Proceedings of the International Computer Music Conference 1999, Beijing}, 1999.

\bibitem{mcvicar2014automatic}
M.~McVicar, R.~Santos-Rodr{\'\i}guez, Y.~Ni, and T.~De~Bie, ``Automatic chord estimation from audio: A review of the state of the art,'' \emph{IEEE/ACM Transactions on Audio, Speech, and Language Processing}, vol.~22, no.~2, pp. 556--575, 2014.

\bibitem{humphrey2012learning}
E.~J. Humphrey, T.~Cho, and J.~P. Bello, ``Learning a robust tonnetz-space transform for automatic chord recognition,'' in \emph{2012 IEEE International Conference on Acoustics, Speech and Signal Processing (ICASSP)}.\hskip 1em plus 0.5em minus 0.4em\relax IEEE, 2012, pp. 453--456.

\bibitem{khadkevich2013reassigned}
M.~Khadkevich and M.~Omologo, ``Reassigned spectrum-based feature extraction for gmm-based automatic chord recognition,'' \emph{EURASIP Journal on Audio, Speech, and Music Processing}, vol. 2013, pp. 1--12, 2013.

\bibitem{ni2012end}
Y.~Ni, M.~McVicar, R.~Santos-Rodriguez, and T.~De~Bie, ``An end-to-end machine learning system for harmonic analysis of music,'' \emph{IEEE Transactions on Audio, Speech, and Language Processing}, vol.~20, no.~6, pp. 1771--1783, 2012.

\bibitem{mauch2010approximate}
M.~Mauch and S.~Dixon, ``Approximate note transcription for the improved identification of difficult chords.'' in \emph{ISMIR}, 2010, pp. 135--140.

\bibitem{lanz2021automatic}
V.~Lanz, ``Automatic chord recognition in audio recording,'' Ph.D. dissertation, Univerzita Karlova, Matematicko-fyzik{\'a}ln{\'\i} fakulta, 2021.

\bibitem{park2019bi}
J.~Park, K.~Choi, S.~Jeon, D.~Kim, and J.~Park, ``A bi-directional transformer for musical chord recognition,'' \emph{arXiv preprint arXiv:1907.02698}, 2019.

\bibitem{wu2020semi}
Y.~Wu, T.~Carsault, E.~Nakamura, and K.~Yoshii, ``Semi-supervised neural chord estimation based on a variational autoencoder with latent chord labels and features,'' \emph{IEEE/ACM Transactions on Audio, Speech, and Language Processing}, vol.~28, pp. 2956--2966, 2020.

\bibitem{li2024large}
C.~Li, J.~Jiang, Y.~Li, and L.~Tian, ``Large-vocabulary chord recognition based on contrastive learning and noisy student (2023),'' \emph{IEEE Transactions on Consumer Electronics}, 2024.

\bibitem{Brown91}
J.~C. Brown, ``Calculation of a constant {Q} spectral transform,'' \emph{The Journal of the Acoustical Society of America}, vol.~89, no.~1, pp. 425–--434, 1991.

\bibitem{zhang2020transformer}
Q.~Zhang, H.~Lu, H.~Sak, A.~Tripathi, E.~McDermott, S.~Koo, and S.~Kumar, ``Transformer transducer: A streamable speech recognition model with transformer encoders and rnn-t loss,'' in \emph{ICASSP 2020-2020 IEEE International Conference on Acoustics, Speech and Signal Processing (ICASSP)}.\hskip 1em plus 0.5em minus 0.4em\relax IEEE, 2020, pp. 7829--7833.

\bibitem{karita2019comparative}
S.~Karita, N.~Chen, T.~Hayashi, T.~Hori, H.~Inaguma, Z.~Jiang, M.~Someki, N.~E.~Y. Soplin, R.~Yamamoto, X.~Wang, \emph{et~al.}, ``A comparative study on transformer vs rnn in speech applications,'' in \emph{2019 IEEE automatic speech recognition and understanding workshop (ASRU)}.\hskip 1em plus 0.5em minus 0.4em\relax IEEE, 2019, pp. 449--456.

\bibitem{dai2019transformer}
Z.~Dai, ``Transformer-xl: Attentive language models beyond a fixed-length context,'' \emph{arXiv preprint arXiv:1901.02860}, 2019.

\bibitem{wang2019learning}
Q.~Wang, B.~Li, T.~Xiao, J.~Zhu, C.~Li, D.~F. Wong, and L.~S. Chao, ``Learning deep transformer models for machine translation,'' \emph{arXiv preprint arXiv:1906.01787}, 2019.

\bibitem{nguyen2019transformers}
T.~Q. Nguyen and J.~Salazar, ``Transformers without tears: Improving the normalization of self-attention,'' \emph{arXiv preprint arXiv:1910.05895}, 2019.

\bibitem{wu2020lite}
Z.~Wu, Z.~Liu, J.~Lin, Y.~Lin, and S.~Han, ``Lite transformer with long-short range attention,'' \emph{arXiv preprint arXiv:2004.11886}, 2020.

\bibitem{dauphin2017language}
Y.~N. Dauphin, A.~Fan, M.~Auli, and D.~Grangier, ``Language modeling with gated convolutional networks,'' in \emph{International conference on machine learning}.\hskip 1em plus 0.5em minus 0.4em\relax PMLR, 2017, pp. 933--941.

\bibitem{dong2018speech}
L.~Dong, S.~Xu, and B.~Xu, ``Speech-transformer: a no-recurrence sequence-to-sequence model for speech recognition,'' in \emph{2018 IEEE international conference on acoustics, speech and signal processing (ICASSP)}.\hskip 1em plus 0.5em minus 0.4em\relax IEEE, 2018, pp. 5884--5888.

\bibitem{ramachandran2017searching}
P.~Ramachandran, B.~Zoph, and Q.~V. Le, ``Searching for activation functions,'' \emph{arXiv preprint arXiv:1710.05941}, 2017.

\bibitem{lu2019understanding}
Y.~Lu, Z.~Li, D.~He, Z.~Sun, B.~Dong, T.~Qin, L.~Wang, and T.-Y. Liu, ``Understanding and improving transformer from a multi-particle dynamic system point of view,'' \emph{arXiv preprint arXiv:1906.02762}, 2019.

\bibitem{raffel2014mir_eval}
C.~Raffel, B.~McFee, E.~J. Humphrey, J.~Salamon, O.~Nieto, D.~Liang, D.~P. Ellis, and C.~C. Raffel, ``Mir\_eval: A transparent implementation of common mir metrics.'' in \emph{ISMIR}, vol.~10, 2014, p. 2014.

\bibitem{pauwels2013evaluating}
J.~Pauwels and G.~Peeters, ``Evaluating automatically estimated chord sequences,'' in \emph{2013 IEEE International Conference on Acoustics, Speech and Signal Processing}.\hskip 1em plus 0.5em minus 0.4em\relax IEEE, 2013, pp. 749--753.

\end{thebibliography}

\begin{IEEEbiography}[{\includegraphics[width=1in,height=1.25in,clip,keepaspectratio]{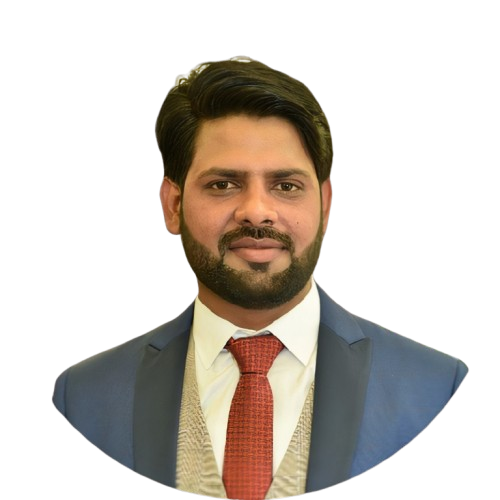}}]{Muhamamd Waseem Akram} received the bachelor's degree in Computer Science from the 
National College of Business Administration \& Economics (NCBAE), Pakistan, in 2014, 
and the Master's degree in Computer Science from COMSATS University Islamabad, Pakistan, in 2019. 
He is currently pursuing the Ph.D. degree in Emerging Digital Technology at 
Scuola Superiore Sant'Anna, Pisa, Italy. His current research focuses on music and artificial intelligence, as well as the application of artificial intelligence in agriculture and steel plant manufacturing.

\end{IEEEbiography}
\begin{IEEEbiography}[{\includegraphics[width=1in,height=1.25in,clip,keepaspectratio]{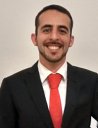}}]{Stefano Dettori} received his master's degree in Automation and Robotics Engineering from the University of Pisa, and his Ph.D in Emerging Digital Technologies from Scuola Superiore Sant'Anna in Pisa. Currently, He is an assistant professor at ICT-COISP (TeCIP Institute), Scuola Superiore Sant'Anna. His research interests include time series forecasting, reservoir computing, artificial intelligence applications, model predictive control and optimization, and design of decision support systems for industrial systems.
\end{IEEEbiography}
\begin{IEEEbiography}[{\includegraphics[width=1in,height=1.25in,clip,keepaspectratio]{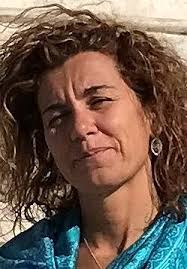}}]{Valentina Colla} (Fellow, IEEE) received her Master degree in Telecommunication Engineering from University of Pisa in 1994 and her Ph.D in Industrial and Information Engineering in 1998 from Scuola Superiore Sant’Anna. She is currently associate professor at Scuola Superiore Sant’Anna. Her research interests cover Artificial Intelligence-based solutions for modelling, simulation, monitoring, and optimization tasks in industrial contexts, with a particular focus on metallurgical and manufacturing industry. In these fields, she authored over 400 papers and participated to 83 projects funded by the European Union, 14 with the role of project coordinator.

\end{IEEEbiography}
\begin{IEEEbiography}[{\includegraphics[width=1in,height=1.25in,clip,keepaspectratio]{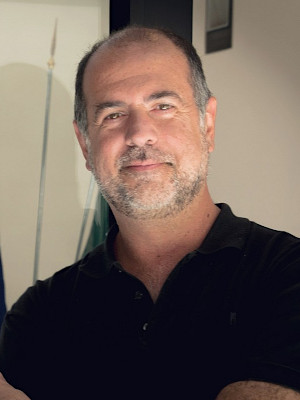}}]{Giorgio Carlo Buttazzo} (Fellow, IEEE) received the
degree in electronic engineering from the University of Pisa, the M.S. degree in computer science
from the University of Pennsylvania, and the Ph.D.
degree in computer engineering from the Scuola
Superiore Sant’Anna, Pisa. He is currently a Full
Professor of computer engineering with the Scuola
Superiore Sant’Anna. He has authored seven books
on real-time systems and more than 300 articles in
the field of real-time systems, robotics, and neural networks. He received 13 best paper awards. He has been the Editor-in-Chief of Real-Time Systems and is an Associate Editor of
ACM Transactions on Cyber-Physical Systems.

\end{IEEEbiography}

\vfill

\end{document}